\magnification=1200
\hsize 6.0truein
\vsize 8.5truein
\parindent 1.0truecm
\font\smallrm=cmr8			
\def\gs{\buildrel > \over \sim}
\def\ls{\buildrel < \over \sim}


\rm
\null


\footline={\hfil}
\vglue 0.8truecm

\rightline{ hep--ph/9512440 }
\rightline{\bf DFUPG--95--GEN--01 }
\rightline{\sl December 1995 }
\vglue 2.0truecm

\centerline{\bf In Search of the Quark Spins in the Nucleon: }
\centerline {\bf A Next--to--Next--to--Leading Order QCD 
Analysis }
\centerline {\bf of the Ellis--Jaffe Sum Rule$^\dagger$ }
\vglue 2.0truecm

\centerline{ Paolo M. Gensini}
\centerline{\sl Dip. di Fisica dell'Universit\`a di Perugia, Perugia, 
Italy, and }
\centerline{\sl Sezione di Perugia dell'I.N.F.N., Perugia, Italy}
\vglue 3.0truecm

\centerline { To be submitted to {\sl Zeitschrift f\"{u}r Physik C: 
Particles and Fields} }
\vglue 6.0truecm

\hrule
\vglue 0.3truecm

\noindent \dag ) Work supported through Fondi 40 \% by the Italian 
Ministry of University and Scientific Research.
\pageno=0
\vfill
\eject



\footline={\hss\tenrm\folio\hss}	
\centerline{\bf In Search of the Quark Spins in the Nucleon: }
\centerline {\bf A Next--to--Next--to--Leading Order QCD Analysis }
\centerline {\bf of the Ellis--Jaffe Sum Rule }
\vglue 1.0truecm

\centerline{ Paolo M. Gensini}
\centerline{\sl Dip. di Fisica dell'Universit\`a di Perugia, Perugia, 
Italy, and }
\centerline{\sl Sezione di Perugia dell'I.N.F.N., Perugia, Italy}

\vglue 1.0truecm
\centerline{\bf ABSTRACT }
\vglue 0.5truecm

{\narrower\smallskip\smallrm
\par The data from the last seven experiments performed on polarized 
deep--inelastic scattering on proton and neutron (or deuteron) targets 
have been analyzed in search of a precise determination of the spin 
fraction carried by the quarks in the nucleon. We find that this 
fraction can be of the size expected from na\"{\i}ve quark model 
arguments, provided the gluon axial anomaly is explicitly included 
and the isosinglet axial charge normalization is fixed at a suitably 
low momentum scale, such that a) the running, strong coupling constant 
is about unit, and b) the orbital angular momentum inside the nucleon 
vanishes.
\par We also find that, despite the appeal of this solution of the 
``nucleon spin crisis'', a solution where the axial anomaly is absent 
and its effects are traded for an appreciable strange quark polarization 
can not however be excluded --- because of the limited accuracy of the 
data --- unless this latter and/or the gluon polarization in the nucleon 
are explicitly measured.
\smallskip}
\vglue 1.0truecm

\leftline{\bf 1. Introduction. }
\vglue 0.5truecm

\par Just after the 1988 publication by the European Muon Collaboration 
(EMC) at CERN of their preliminary$^1$ results on the asymmetry in muon 
polarized deep--inelastic scattering (PDIS) on a polarized hydrogen 
(actually spin--frozen ammonia) target, the particle physics community 
was somewhat shocked to learn that the total sum of the quark spins in 
the proton seemed to be very close to zero, rather than somewhere between 
1/2 and 3/4, as generally expected from na\"{\i}ve quark--model arguments.

\par As for any unexpected discovery in our field, to begin perhaps with 
that of the muon, the following years have seen both the planning and 
running of new experiments, and the deepening of the theoretical studies 
(not completely free of much heated controversies) trying to clarify this 
mystery, often know under the name of ``nucleon spin crisis''.

\par The aim of this paper, relying heavily on the results thus 
accumulated, is to use i) all PDIS data taken on both proton and neutron 
(or deuteron) targets, ii) current phenomenological ideas on the 
behaviour of the parton distributions at both large and small values of 
the Bjorken variable $x$, and iii) the results from perturbative QCD 
(PQCD) at next--to--next--to--leading order (NNLO), to produce an 
internally consistent estimate of the quark spin content of the nucleon.

\par It will turn out that one can not, at the present level of 
experimental accuracy, unambiguously separate the different flavour 
components from PDIS data {\sl alone}, nor directly verify the PQCD 
predictions for the isoscalar sum rule: namely, one can not decide on 
a purely experimental basis on the nature of its unitary--singlet 
component, given the (basically two) different possibilities offered 
{\sl in principle} by the stage at which one decides to send the quark 
masses (used as regulators in the calculations of the splitting 
functions) to zero.

\par Despite this persistent lack of {\sl experimental} proof of all PQCD 
arguments, it can however be shown that, for the ``most natural'' (in the 
parton model framework) of these two possibilities, one can interpret 
the data as being consistent with a negligible intrinsic strange 
component of the quarks' spins at a suitably low mass scale, such that 
$\alpha_s \simeq 1$, consistent with the na\"{\i}ve quark model expectation.

\par There are two primary reasons why this result contradicts the 
initial findings from the EMC data$^1$: the first, of experimental nature, 
is that it is now clear that the $g_1^p$ values 
given by the EMC were too low because of their choice of the 
$F_2^p$ data employed to normalize them; the second, 
of theoretical nature, is that PQCD corrections were included only at leading 
order, while it is only at next--to--leading order that peculiar features of 
the {\sl polarized} deep--inelastic scattering (and particularly for its 
unitary--singlet piece) become to emerge, as it will be clear in next section.

\par The rest of this paper will be divided in three parts: a presentation of 
the PQCD NNLO corrections to the first--moment isovector and isoscalar sum 
rules, a careful enumeration of the constraints that can be imposed on the 
various polarized distribution functions of the quarks, in terms of 
which one can describe the polarized structure functions $g_1^{p,n}$, and 
then a short discussion on the results obtained fitting to the data simple 
parametrizations satisfying these constraints. This discussion will be 
centred on the spin composition of the nucleon, 
and in particular on the need (or absence of any need) for an intrinsic 
strange component $\Delta s$ in it.

\vglue 1.0truecm
\leftline{\bf 2. The First Moments of $g_1^{p,n}$ and PQCD at NNLO. }
\vglue 0.5truecm

\par The sum rules for the first moments $I_0^{p,n}(Q^2) = \int_0^1 dx 
g_1^{p,n}(x,Q^2)$ are often discussed separating the non--singlet, $(p - n)$ 
combination from the isoscalar one, $(p + n)$, which contains both a 
non--singlet and a singlet part. Actually, despite the technical difficulties 
involved in dealing with the latter, both sum rules have equal footing in 
PQCD and should not be considered as separate entities, apart from historical, 
or practical, considerations.

\par Their only difference lies indeed in the fact that the axial coupling 
involved in the first is extremely well measured from neutron $\beta$--decay, 
while the couplings involved in the second are outside direct measurement, 
and can be arrived at only through more or less founded theoretical arguments. 
It is therefore expedient to separate them, since the first can either 
be considered a good test of higher--order PQCD, or, alternatively (and this 
will be the attitude taken in the present paper), a useful normalization 
for the non--singlet part of the moments $I_0^{p,n}(Q^2)$.

\par The first--moment sum rule for the difference between proton and neutron 
polarized structure functions, known as the Bjorken sum rule (BjSR), reads$^2$ 
(written in its full QCD garb)
$$
I_0^{p-n}(Q^2) = \int_0^1 dx [ g_1^p(x,Q^2) - g_1^n(x,Q^2) ] = {1\over6} 
\cdot C_8(\alpha_s) \cdot g_A + (h.t.)_{I_t=1} \ ,
\eqno (1)
$$
where the coefficient $C_8(\alpha_s)$ includes the PQCD corrections to the 
parton--model result, and the symbols $(h.t.)_{I_t=0,1}$ stand for the 
higher--twist contributions$^3$ (HTC), behaving as integer inverse powers 
of $Q^2$. Since experiments extend from $<Q^2> =$ 2.0~GeV$^2$ (SLAC E142 
Collaboration$^4$) to the 10.7~GeV$^2$ of the EMC$^{1,5}$, inclusion of 
these effects turns out to be crucial to a test of BjSR, Eq. (1). This 
is indeed also the attitude taken by Ellis and Karliner$^6$ 
in their analisys of PDIS data. HTC are also essential$^7$ to connect the 
PDIS sum rules for proton and neutron targets to the Gerasimov--Drell--Hearn 
sum rules$^8$ at $Q^2=0$.

\par An estimate of the HTC was originally given --- and revised --- by 
Balitski\u\i, Braun and Kolesnichenko$^3$, and then repeated by several 
groups of authors$^{9,10}$: in this paper preference will be given to the 
estimate by Ross and Roberts$^9$, mostly because of its widespread use 
in other analyses of the same data.

\par Let us now focus our attention on some aspects of the recently 
improved$^{11}$ evaluation of $C_8(\alpha_s)$ and on their implications. 
This coefficient is now known to better than $O(\alpha_s^3)$, and to NNLO 
accuracy it can be written as
$$
C_8(\alpha_s) = 1 - {\alpha_s\over\pi} \cdot [ 1 + c^{(8)}_1(N_f) \cdot 
{\alpha_s\over\pi} + c^{(8)}_2(N_f) \cdot ({\alpha_s\over\pi})^2 + ... ]\ ,
\eqno (2)
$$
where the numerical values of $c^{(8)}_{1,2}$ are listed, for $N_f$ from 3 
to 5, in Table I, and their complete expressions can be found in the original 
paper by Larin and Vermaseren$^{11}$. Due to the large values of the 
constants $c^{(8)}_{1,2}$ for $N_f = 3$, 4, $C_8(\alpha_s)$ is evidently 
decreasing with $\alpha_s$ much more steeply than expected on the basis of the 
leading order estimate $C_8 = 1 - \alpha_s/\pi$ at low values of $Q^2$, 
such as those of SLAC experiments E142$^4$ ($<Q^2> =$ 2.0~GeV$^2$) and 
E143$^{12,13}$ ($<Q^2> =$ 3.0~GeV$^2$), and of the preliminary deuteron data 
from the Spin Muon Collaboration$^{14}$ (SMC) at CERN ($<Q^2> =$ 4.6~GeV$^2$).

\par An initial comment is in order here: actually, the expansion in 
$\alpha_s$ for $C_8(\alpha_s)$ is known to one order 
beyond$^{15,16}$ what has been written in Eq. (2), but its unitary--singlet 
partner, $C_1(\alpha_s)$, has been {\sl estimated}$^{16}$ only to 
$O(\alpha_s^3)$, as most of the quantities computed in PQCD to extract 
the coupling $\alpha_s$: for its consistency, any PQCD analysis {\sl must} 
be performed to {\sl the same, fixed order} in the coupling, using the 
$\beta$--function {\sl at that order} to express the running of 
the coupling $\alpha_s(Q^2)$.

\par Accordingly, here {\sl only} NNLO expansions will be used, together 
with the expansion for $\alpha_s(Q^2)$ (and the estimate of its scale 
$\Lambda_{\overline{MS}}(N_f)$) at the same order: 
a recent PQCD analysis by Bethke and 
Catani$^{17}$ of {\sl all} available data (both space-- and time--like, 
but with no inclusion of PDIS ones) led to a {\sl conservative} estimate of 
$\Lambda_{\overline{MS}}(5)$ = 200~$\pm$~50~MeV, and therefore to the value 
$\Lambda_{\overline{MS}}(3)$ = 410~$\pm$~100~MeV that will be employed 
throughout. There is a reason to overlook more recent compilations: since 
these tend to include also $\alpha_s$ determinations from PDIS data to 
enrich the available points at low momenta, one should exclude these latter 
determinations from the analysis$^{18}$ to be free of possible internal 
biases, and to treat PQCD corrections as external inputs rather than as 
parameters to be fit to the data. This is simply and easily 
accomplished, rather than by re--doing all fits, by using the 
$\Lambda_{\overline{MS}}(3)$ from the slightly older compilation of Ref. 17.

\vglue 0.6truecm
\centerline{\bf Table I}
\centerline{\bf Coefficients of higher QCD corrections to BjSR }
\vglue 0.3truecm
\hrule
\vglue 0.3truecm
$$\vbox{\halign{# \hfil & \qquad \hfil # \hfil & \qquad \hfil # \hfil \cr
$N_f$ & $c_1(N_f)$ & $c_2(N_f)$ \cr
3 & 3.5833 & 20.2153 \cr
4 & 3.2500 & 13.8503 \cr
5 & 2.9167 & 7.8402 \cr}}$$
\hrule
\vglue 0.6truecm

\par A question better addressed at this point is the actual value of $N_f$, 
the number of active flavours, to be used at each $Q^2$ in the PQCD 
expansions of the coefficients $C_{1,8}(\alpha_s)$ 
and of the anomalous dimension 
of the unitary--siglet axial charge. For time--like $Q^2$, there is no 
ambiguity, since for heavy quarks the flavour thresholds at $Q^2=4m_i^2$ 
can be fixed by setting the mass $m_i$ of the $i$--th flavour quark 
approximately equal to that of its lowest--mass pseudoscalar meson. In 
deep--inelastic scattering, however, a flavour is active {\sl only when 
appreciably contributing} to the moment sum rules, {\sl i.e.} when 
produced a) in a {\sl really inclusive} manner (read: not only in 
low--multiplicity events), and b) over an {\sl appreciable range} in 
Bjorken's variable $x$ (say up to $x \simeq 1/3$). If one sets the 
beginning of the scaling region at $Q^2 \simeq$ 2~GeV$^2$ (as indicated 
by the ``classic'' SLAC--MIT experiments), the previous requirements ask 
for a $Q^2 \gs$ 17~GeV$^2$ for charm to be an active flavour in DIS, 
and so mandates $N_f =$ 3 for the PQCD analysis of all available PDIS data.

\par The choice of $N_f =$ 3 (rather than 4) has no great effect on 
$C_8(\alpha_s)$, as one can read from Table I, but for the 
$Q^2$--evolution of the unitary--singlet piece the coefficient of 
the first--order term in the expansion in powers of $\alpha_s$ of 
its integrated anomalous dimension almost cancels, for $N_f = 4$, 
the first--order one in $C_1(\alpha_s)$. It is therefore important, 
for the extraction of the quarks' spin fraction, 
$$
\Sigma = \sum_i^{N_f} \Delta q_i = \Delta u + \Delta d + \Delta s \ \ \ 
{\rm for}\ N_f = 3 \ ,
\eqno (3)
$$
to use the value of $N_f$ appropriate to the range of values of $Q^2$ 
where the actual data have been taken. This, unfortunately, has not 
always been done consistently by some authors$^{19}$.

\par In 1974, Ellis and Jaffe$^{20}$, faced with the problem of how to 
use a sum rule akin to Eq. (1) with only hydrogen data, used parton--model 
ideas, flavour SU(3) symmetry and the Okubo--Zweig (OZI) rule 
to derive a sum rule for the 
first moment of $g_1^p$ {\sl alone}; as stated above, the PQCD--corrected 
version of such a sum rule (when freed of OZI--rule restrictions) is a 
part of PQCD as fundamental as the BjSR. For the isoscalar combination 
of PDIS structure functions this sum rule becomes the (PQCD corrected) 
Ellis--Jaffe sum rule (EJSR) 
$$
I_0^{p+n}(Q^2) = \int_0^1 dx [ g_1^p(x,Q^2) + g_1^n(x,Q^2) ] = 
$$
$$
= {1\over{18}} \cdot C_8(\alpha_s) \cdot g_8 + {2\over9} \cdot C_1(\alpha_s) 
\cdot g_0(Q^2) + (h.t.)_{I_t=0} \ ,
\eqno (4)
$$
not to be confused with the {\sl original} EJSR$^{20}$, which was 
derived for $I_0^p(Q^2)$ {\sl only}, and without any PQCD 
corrections to the parton--model plus OZI--rule predictions.

\par Additional complications with respect to the BjSR, Eq. (1), arise from 
the facts a) that the isoscalar axial charges of the nucleon are not 
directly measurable, and $g_8$ is indeed derived via flavour--symmetry 
arguments (apart from symmetry--breaking effects), and b) that the 
unitary--singlet one $g_0(Q^2)$ couples not only to the quarks, but also 
to the gluons via the axial anomaly and possesses therefore anomalous 
dimensions$^{21}$, so that its evolution with $Q^2$ is not exausted by 
the PQCD coefficient $C_1(\alpha_s)$ and must be explicitly computed.

\par Since this coupling is scheme dependent, this point has been the focus 
of a very heated theoretical debate$^{22}$. To cut a long history short, one 
can summarise it by saying that, in conventional parton language where the 
masses of the partons $m_i$ are neglected with respect to the momentum 
scale $Q$, and wishing to identify $\Sigma$ with the spin fraction carried by 
the quarks {\sl actually} present in the target proton, one has to put
$$
g_0(Q^2) = \Sigma - N_f {\alpha_s\over{2\pi}} \Delta G(Q^2) \ ,
\eqno (5)
$$
where $\Delta G$ is the first moment of the gluon polarized distribution 
function $\delta G(x) = G_+(x) - G_-(x)$, and determine its evolution 
via the equation (where $t = \log Q^2/\mu^2$) 
$$
{d\over{dt}} g_0(t) = - N_f {\alpha_s\over{2\pi}} \gamma_{gq}(\alpha_s) g_0(t) 
\ ,
\eqno (6)
$$
which relates to the anomalous dimension of the axial anomaly via 
$$
- N_f {\alpha_s\over{2\pi}} \gamma_{gq}(\alpha_s) = \gamma_{gg}(\alpha_s) - 
\beta(\alpha_s) {{2\pi}\over\alpha_s} 
\eqno (6')
$$
where the various $\gamma_{ij}$ ($\{i,j\} = \{g,q\}$) represent the 
coefficients giving $d\Sigma/dt$ and $d\Delta G/dt$ in terms of $\Sigma$ 
and $\Delta G$, whose matrix is diagonalized (in any scheme where quark 
mass regulators are sent to zero) building the combination in Eq. (5) on one 
side, which evolves anomalously as in Eq. (6), and leaving $\Sigma$ on 
the other, free of anomalous dimensions since the constraint in Eq. (6$'$) 
makes the determinant of the $2 \times 2$ matrix of the (final) coefficients 
(whose eigenvalues are the anomalous dimensions of the two operators mixed 
by the evolution) vanish identically. This has been verified step by step 
at next--to--leading order$^{22}$: that it should hold also at higher orders 
is inferred from the fact that conserved (and partially conserved) charges 
should be free of anomalous dimensions on one side, and on the other there 
is nothing, apart from the mixing with the unitary singlet of the pure 
gluonic world, to distinguish an SU(3)-- from a U(3)--symmetric fermionic 
world, so that it is ``natural'' to expect {\sl all} purely fermionic 
operators to be free of anomalous dimensions.

\par After integrating in $\alpha_s$ from a {\sl normalization} scale 
$\mu^2$ to $Q^2$, one obtains the integrated anomalous dimension at NNLO 
$$
\log\ {{g_0(Q^2)}\over{g_0(\mu^2)}} = \tilde\gamma(\alpha_s(Q^2)) - 
\tilde\gamma(\alpha_s(\mu^2)) = {{6N_f}\over{33-2N_f}} 
{{\alpha_s(Q^2)-\alpha_s(\mu^2)}\over\pi} \cdot 
$$
$$
\cdot [\ 1 + \ ({{83}\over{24}} + {N_f\over{36}} - 
{{33-2N_f}\over{8(153-19N_f)}}) \ 
{{\alpha_s(Q^2)+\alpha_s(\mu^2)}\over\pi} + ...\ ] \ ,
\eqno (7)
$$
with the NNLO calculation by Larin$^{23}$ of the anomalous dimension.

\par As one can read from the above expression, $g_0(Q^2)$ can be drastically 
reduced (still at $N_f = 3$) from its value at the {\sl normalization} scale 
$\mu^2$ for $Q^2 \gg \mu^2$; on the other hand, one can not set 
$\mu^2 \to \infty$ and thus drop $\alpha_s(\mu^2)$ 
from Eq. (7) for two reasons: 
first, that the anomaly contribution to Eq. (5) is not definable in this 
limit$^{21,22}$, due to $\Delta G(Q^2)$ diverging asymptotically as 
$\alpha_s(Q^2)^{-1}$ from Eqs. (5) and (7), and, second, that we can not 
keep $N_f = 3$ while letting $Q^2 \to \infty$.

\par There is another point which deserves consideration: the expression in 
Eq.~(7) is of second order in $\alpha_s$, while it was said from the beginning 
that the analysis of the PDIS data will be done {\sl consistently} at NNLO, or 
$O(\alpha_s^3)$. Indeed the anomalous dimension $\gamma_{gg}(\alpha_s)$ has 
been computed to $O(\alpha_s^3)$ by Larin$^{23}$: it is however a logarithmic 
derivative in the variable $t$, which to be integrated has to be divided 
by the $\beta$--function to make the variable change from $t$ to $\alpha_s$, 
losing in this step a power in the coupling $\alpha_s$. This gives no 
problem if the anomalous dimension is left where it belongs, {\sl i.e.} as 
the argument of an exponential function, $g_0(Q^2) = g_0(\mu^2) \cdot \exp 
[\tilde\gamma(\alpha_s(Q^2)) - \tilde\gamma(\alpha_s(\mu^2))]$, where 
$\tilde\gamma(\alpha_s)$ is the {\sl integrated} anomalous dimension, 
but has caused unnecessary problems whenever $\exp\tilde\gamma(\alpha_s)$ 
has been expanded in powers of the coupling, since then one had either 
to go one step further in the expansion of the anomalous dimension to 
find the ``missing'' power$^{15,16}$, or to truncate ``prematurely'' the 
series in the coefficient $C_1(\alpha_s)$. Both procedures are inconsistent 
from a PQCD point of view, since they tend to mix one order with the next, 
while the running coupling $\alpha_s$ must be defined order by order: it is 
clear that in this way one will end up using $\alpha_s$ at a given order 
in at least one perturbative expansion calculated at a different order (not 
to be confused with the power of $\alpha_s$ appearing in the {\sl final} 
expression, which could depend, as is the case here, on additional 
mathematical manipulations).

\par Another, equally unnecessary, but luckily only {\sl semantic} problem 
has been created by the persons who first misnamed the scale $\mu^2$ in the 
same equation a {\sl renormalization} scale, while it is clear enough that 
one has to do with just a normalization scale, which has absolutely nothing 
to share with the actual renormalization procedure$^{15}$. The scale 
$\mu^2$ is thus free both to appear explicitly in the physical expressions, 
and to be chosen to follow the author's (or authors') theoretical prejudices; 
certain precautions must however be followed: for instance, working at a 
{\sl fixed} number of flavours $N_f = 3$, it would be rather unwise to set 
it to infinity, where $N_f$ will be six at the best of our knowledge, so 
that the $g_0(\infty)^{N_f=3}$ so obtained would hardly be connected to the 
{\sl physical} limit of $g_0(Q^2)$ for $Q^2 \to \infty$. Note also that this 
$g_0(\infty)^{N_f=3}$ (whatever its meaning) {\sl can not} be identified 
with $\Sigma$ as defined by Eq. (3): if one believed the singlet axial 
coupling $g_0(Q^2)$ in the EJSR, Eq. (4), to be given by $\Sigma$ {\sl 
alone}, from the diagonalization of the unitary--singlet operators one 
had also to put $\tilde\gamma = 0$ accordingly. In line of principle, 
these two definitions for $g_0(Q^2)$ 
could be distinguished on the basis of the different evolutions 
with $Q^2$ they predict for the EJSR integrals $I_0^{p+n}(Q^2)$: 
making the two to coincide at a scale $Q_0^2$, 
the difference for the EJSR between the case of a running $g_0(Q^2)$ 
and that of $g_0 = \Sigma$ at a scale $Q^2$ would correspond to 
$$
\Delta I_0^{p+n}(Q^2) = {2\over9} \cdot C_1(\alpha_s(Q^2)) \cdot \{ 
\exp [\tilde \gamma (\alpha_s(Q^2)) - \tilde \gamma (\alpha_s(Q_0^2)) ] - 1 \} 
\cdot g_0(Q_0^2)\ .
\eqno (8)
$$
We shall return later on this point, when analysing our numerical results in 
Section 4.

\par The evaluation of the singlet coefficient $C_1(\alpha_s)$ runs along the 
same lines as that of the non--singlet one $C_8(\alpha_s)$ and it can be 
similarly expanded in powers of $\alpha_s$: 
$$
C_1(\alpha_s) = 1 - {\alpha_s\over\pi} \cdot [ 1 + c^{(1)}_1(N_f) \cdot 
{\alpha_s\over\pi} + c^{(1)}_2(N_f) \cdot ({\alpha_s\over\pi})^2 + ... ]\ .
\eqno (9)
$$
There are hovewer additional complications, for this coefficient receives 
contributions from graphs, not included in the other, whose number grows 
with the order of the calculation. Thus Kataev$^{16}$ has produced only 
an estimate of the constant $c^{(1)}_2$ for $N_f = 3$ (note that one of 
the conditions under which this was possible breaks down for $N_f = 4$).

\par To close this section, the numerical values of the constants for 
the perturbative expansions in orders of $(\alpha_s/\pi)$ of $C_8(\alpha_s)$, 
$C_1(\alpha_s)$, and $\tilde\gamma(\alpha_s) = \sum_k \gamma_k \cdot 
(\alpha_s/\pi)^k$, are listed in Table II. Two points must be noted 
before going to the next section: the constants in the singlet part are 
systematically lower than in the corresponding, non--singlet one, 
giving a slower evolution with $Q^2$, and the trends of the integrated 
anomalous dimension $\exp\tilde\gamma(\alpha_s)$ and of the coefficient 
$C_1(\alpha_s)$ run in opposite directions, tending to some extent to 
compensate each other.

\par Also, the first line in Table II makes clear enough the essential 
difference between leading-- and higher--order PQCD treatments: besides 
introducing an anomalous dimension for $g_0(Q^2)$, the latter break strongly 
the {\sl accidental}, lowest--order degeneracy of the coefficient functions 
$C_{1,8}(\alpha_s)$.

\vglue 0.6truecm
\centerline{\bf Table II}
\centerline{\bf Constants in the perturbative expansions at NNLO }
\vglue 0.3truecm
\hrule
\vglue 0.3truecm
$$\vbox{\halign{#\hfil&\qquad\hfil#\hfil&\qquad\hfil#\hfil&\qquad\hfil#\hfil
\cr
$k$ & $c_k^{(8)}$ & $c_k^{(1)}$ & $\gamma_k$ \cr
0 & 1 & 1 & 0 \cr
1 & 3.5833 & 1.0959 & 0.6667 \cr
2 & 20.2153 & $\sim$ 3.7 & 0.9907 \cr}}$$
\hrule
\vglue 0.6truecm

\vglue 1.0truecm
\leftline{\bf 3. Measurements and parametrizations for $g_1(x)$. }
\vglue 0.5truecm

\par What is actually measured by experiments in PDIS are not the structure 
functions $g_1$ themselves, but rather the polarization asymmetries $A_1$, 
related to the polarized cross sections $\sigma^{\uparrow\uparrow}, 
\sigma^{\uparrow\downarrow}$ by
$$
D \cdot A_1 = (\sigma^{\uparrow\uparrow} - \sigma^{\uparrow\downarrow}) / 
(\sigma^{\uparrow\uparrow} + \sigma^{\uparrow\downarrow})\ ,
\eqno (10)
$$
where $D$ is the target polarization fraction, and $g_2$ is neglected: $g_1$ 
is then related to $A_1$ by 
$$
g_1(x,Q^2) = {{A_1 \cdot F_2(x,Q^2)}\over{2x\cdot [1 + R(x,Q^2)]}} \ .
\eqno (11)
$$

\par It is obvious that the factor $2x$ in the denominator makes a direct 
determination of $g_1$ at $x \to 0$ impossible for finite--accuracy data. 
The evaluation of the EJSR, Eq. (4), depends thus on the 
parametrization assumed to extrapolate $g_1$ to $x = 0$: the usual 
treatment of this point has till now assumed it to extrapolate {\sl 
smoothly} to a constant, as $g_1 \sim \alpha + \beta \cdot x$, in accord 
with the pion--pole trajectory intercept being close to zero. However, 
since one does not expect the sea distributions to couple dominantly to 
an isovector, pseudoscalar trajectory, but rather to an isoscalar one 
such as the eta, one should rather have for these a behaviour 
$x^{-\alpha_\eta(0)}$, with $\alpha_\eta(0) \simeq -1/4$, which, together 
with the expected negative sign for the sea contribution, 
produces a spike in the 
{\sl isoscalar part} of $g_1$ as $x \to 0$, of the type $g_1 \sim 
\alpha - \beta \cdot x^{1/4}$ (with $\alpha$, $\beta > 0$), perhaps 
just appearing$^{24}$ at very low values of the Bjorken variable $x$ in the 
proton data taken by the Spin Muon Collaboration at CERN at $<Q^2>=$ 10.0 
GeV$^2$. Of course, such a spiky behaviour 
does not show in the integrand of the BjSR, which can therefore be 
extrapolated smoothly to $x = 0$ according to the conventional practice 
in this matter.

\par This point could have a non--negligible influence 
on the evaluation of the EJSR, Eq. (4), raising the low--$x$ contribution 
to its left--hand side well above the conventional estimates. A special 
comment is in order here on the EMC published data$^5$ for $g_1^p$: 
rather than using them, one should use instead only 
their values for $A_1^p$ with an {\sl adequate} set of values for 
$F_2^p$ and $R^p$. Indeed, the EMC calculated $g_1^p$ using a) 
the ratio $R^p$ predicted by PQCD, systematically smaller than 
experiment since finite--mass corrections are dominant at low 
values of $Q^2$ and $x$ (though the effect of this choice is not 
too important at $<Q^2> = $ 10.7~GeV$^2$), 
and b) {\sl their} values for the {\sl unpolarized} structure function 
$F_2^p$, systematically lower than those by the BCDMS 
collaboration$^{25}$ (and than the recent New Muon Collaboration (NMC) 
data$^{26}$ as well) by as much as 13\% at the lowest values of $x$. 
Even using their {\sl measured} values of $A_1$ together with a 
phenomenological parametrization for $F_2^p(x,Q^2)$ (and $R^p(x,Q^2)$) to 
produce $g_1^p(x,Q^2)$ at a reference, fixed value of $Q^2$ (and assuming 
$A_1$ to vary little$^{6,15,27}$ with $Q^2$, an assumption which a recent 
analysis from the E143 Collaboration at SLAC seem to corroborate$^{28}$) 
is not completely free of the above, last source of error: 
indeed only the latest, post--NMC parametrizations$^{29}$ have dropped 
the {\sl unpolarized} EMC data altogether, while {\sl all} previous 
analyses ended up averaging over the two, conflicting sets of data for 
small values of $x$. We have therefore re--normalized the {\sl published} 
EMC data$^5$ for $g_1^p$ with the known ratio of BCDMS to EMC data, 
and this will be the data set referred to as EMC$^p_{rev}$ in 
the rest of the paper.

\par As just mentioned two paragraphs above, to build the first moments 
the actual measurements of $g_1^{p,n}$ have to be extrapolated in $x$ to 
$x = 0$ and to $x = 1$, to cover parts of the integration range not 
coverable by the experiments on the asymmetry $A_1^{p,n}$. The second 
extrapolation does not pose any problem, since $A_1$ tends to unit (and 
$R$ to zero) in the limit $x \to 1$ (this is a well--known feature$^{30}$ 
of the nucleon wavefunction: at high values of $x$ the proton --- neutron --- 
structure is dominated by the $u_+$ --- $d_+$ --- quark distribution: some 
parametrizations$^{31}$ violate this constraint gaining thus additional but 
unphysical freedom in dealing with the small sea components), 
and Eq. (11) is thus 
reducing simply to $g_1 \sim F_2/(2x)$, whose behaviour as $x \to 1$ is 
largely determined by well known ``counting rules'' of conventional 
parton models.

\par In this paper, instead of extrapolating {\sl separately} in the two 
extreme ranges of $x$, the choice has been made to {\sl fit} the data to 
simple functional forms incorporating at least three elements: a) the 
counting rules, b) reasonable Regge (or other, QCD--motivated) behaviours 
for $x \to 0$ (separately for valence and sea contributions), and c) 
constraints on the integrals of the parton distribution functions coming 
from a connection between the constituent--quark picture and the 
quark--parton model, originally introduced by Altarelli, Cabibbo, Maiani 
and Petronzio$^{32}$, and recently recovered in this context 
by Fritzsch$^{33}$.

\par In this picture, the structure functions $g_1^{p,n}$ are decomposed 
in terms of the helicity distribution fuctions $\delta q_i'$ for each 
active quark flavour ($q_i = u$, $d$, $s$), where the prime indicates that 
the contribution from the axial anomaly, formally of order $\alpha_s$, has 
been included in each flavour's sea distribution, $\delta q_i' = \delta q_i 
+ k_{qg} \otimes \delta G$ (the symbol $\otimes$ will stand from here 
on for a convolution integral in Bjorken variables), with $k_{qg}$ the 
appropriate gluon--to--quark splitting function. It has been sometimes said 
in the literature that the anomaly contributes to $g_1$ only at very small 
$x$ values, so that it should be hardly seen in the $x$--regions covered by 
experiments: this is true (in the scheme where quark masses $m_i$ are 
sent to zero) {\sl only if} the polarized gluon distribution is assumed to 
peak at $x = 0$, as {\sl e.g.} in the intrinsic gluon distribution proposed 
by Brodsky and Schmidt$^{34}$. Unfortunately, their distribution for 
$\delta G(x)$ has the wrong Regge behaviour for $x \sim 0$, {\sl contrary 
to their statement}, since it requires dominance of $\delta G$ by the pion 
trajectory, with intercept $\alpha_\pi(0) \simeq 0$, while it should be 
dominated instead by a pseudoscalar--glueball trajectory, with an expected 
intercept $\alpha_G(0) \ls \alpha_{\eta\prime}(0) \simeq -1$. When this 
constraint is imposed on the Brodsky--Schmidt formul\ae, {\sl ceteris 
paribus}, most of the anomaly contribution falls in the $x$ interval 
covered by the experiments ({\sl e.g.} 80\% of it in the case of the EMC 
$x$--range), making it virtually indistinguishable from the other, intrinsic 
sea distributions, while the integral $\Delta G$ remains of the same 
magnitude as in Ref. 34, being solidly tied to the momentum fraction 
$<x_G> \simeq 1/2$ carried by the gluons at the momentum scale 
$<Q^2>_{SLAC} \simeq$ 1 $\sim$ 2~GeV$^2$ at which these intrinsic components 
are defined.

\par Separating further valence ($u_v$, $d_v$) from sea ($u'_s$, $d'_s$, 
$s'$) components one can write, with self--explanatory notations, 
$$
g_1^p(x,Q^2) = {2\over9} \cdot [\delta u_v(x,Q^2) + \delta u'_s(x,Q^2) ] + 
$$
$$
+ {1\over{18}} \cdot [\delta d_v(x,Q^2) + \delta d'_s(x,Q^2) + \delta 
s'(x,Q^2) ] \ ,
\eqno (12)
$$
and
$$
g_1^n(x,Q^2) = {2\over9} \cdot [\delta d_v(x,Q^2) + \delta d'_s(x,Q^2) ] + 
$$
$$
+ {1\over{18}} \cdot [\delta u_v(x,Q^2) + \delta u'_s(x,Q^2) + \delta 
s'(x,Q^2) ] \ .
\eqno (13)
$$
The valence and sea distributions (with the latter primed to distinguish 
them from the intrinsic ones, free of the gluonic contribution, which 
relate to the quarks spin contents $\Delta q_i$ entering {\sl e.g.} in the 
angular momentum sum rule), will in general be expressed by the general 
functional form $\delta q_{v,s}'(x,Q^2) = \sum_j x^{-\alpha_j(0)} \cdot 
(1-x)^{n_{v,s}} \cdot P_j(x,Q^2)$, where the sum runs over the Regge 
singularities assumed to dominate the distribution at $x \sim 0$, and the 
$P_j(x,Q^2)$ will be polynomials in $x$ with $Q^2$--dependent 
coefficients$^{35,36}$.

\par For the powers $n_{v,s}$ the parton--model counting rules$^{36}$ give 
$n = 2N_s-1$ (where $N_s$ is the number of ``spectator'' quarks), or $n_v 
=$ 3 and $n_s =$ 7 (for gluons the rule gives $n_g =$ 5: their further 
suppression in the distributions $\delta q'_s$ at high $x$--values comes 
from the splitting function): it is hovewer known$^{35,36}$ that the valence 
u--quark dominates at high $x$ values over the d--quark in their 
unpolarized distributions, and this fact is commonly ``explained'' as a 
consequence of the Pauli principle, barring two quarks of the same quantum 
numbers from being close to each other in phase space, and often 
expressed as a 
``penalty factor'' $(1 - x)$ in the odd flavour distribution function. 
The same mechanism should operate in the polarized valence distributions 
as well, as in the sea distributions (both polarized and unpolarized), 
suppressing here u--quarks with respect to d--quarks, the ``penalty'' 
being now paid when an u--antiquark is produced at $ x \sim 1$; while 
there is undisputable evidence of this effect for unpolarized {\sl valence} 
distributions in the ratio $F_2^n/F_2^p$, tending almost linearly 
to the value 1/4 as 
$x \to 1$, the same effect in the unpolarized {\sl sea} ones could be 
responsible$^{37}$ for the defect of the so--called Gottfried sum rule (one 
can simply check that the figures are indeed of the right order of 
magnitude, though a detailed model would require a complete re--fitting 
of all unpolarized parton distributions). For the PDIS data, this 
Pauli--principle ``penalty factor'' will be included only in the 
polarized {\sl valence} distributions, for the effects of its presence in the 
sea ones would be so small {\sl vis--\`a--vis} the experimental errors 
that its inclusion would only lead to unnecessary mathematical 
complications in the fitting procedures.

\par With this additional factor omitted, and reducing the polynomials 
$P_j(x,Q^2)$ to $Q^2$--dependent factors, independent of $x$, the sea 
contributions to the PDIS structure functions will reduce to an 
isoscalar term, simply expressed as
$$
g_1^{p,n}(x,Q^2)_{sea} = P(Q^2) \cdot x^{1\over4} \cdot (1 - x)^7 \ ,
\eqno (14)
$$
where we have assumed all sea components to couple dominantly to the 
$\eta$--meson trajectory with intecept $\alpha_\eta(0) \simeq - 1/4$. 
Note that under this hypothesis the sea would contribute to 
the BjSR only through the difference 
between u--quark and d--quark seas eventually brought about by the 
Pauli--principle ``penalty factor'': once this latter is dropped, the 
BjSR contains only the valence distributions and is therefore offering 
a mean to define their normalizations in terms of $g_A$.

\par This is accomplished by using the connection between constituent--quark 
and parton model languages of Refs. 32--33, {\sl i.e.} by writing $\delta u = 
\delta U \otimes \delta u_U + \delta D \otimes \delta u_D$ (and so on for all 
active flavours and for gluons as well), where the notation $\delta q_Q$ 
represent the $q$--flavoured parton polarized density inside the 
$Q$--flavoured constituent quark. Assuming SU(2) symmetry to hold for the 
partons inside the constituent quarks ($Q = U$, $D$ only), and integrating 
over all Bjorken variables, one can put $\Delta u_U = \Delta d_D = \Delta 
(q_Q)_v + \Delta (q_Q)_s$, $\Delta d_U = \Delta u_D = \Delta (q_{Q'})_s$ and 
$\Delta s_U = \Delta s_D = \Delta s$, and one finds the relation 
for the spin content of the valence partons
$$
{{\Delta u_v}\over{\Delta d_v}} = {{\Delta U}\over{\Delta D}} = - 4
\eqno (15)
$$
from the constituent quark model results $\Delta U = 4/3$, $\Delta D = 
-1/3$: note that, since these constituent quarks are structured objects, 
these values {\sl do not} imply $g_A = 5/3$ ({\sl modulo} small recoil 
corrections), as in na\"{\i}ve treatments of the constituent quark model. 
The same hypothesis leads only to the generous bounds $- 1/4 < \Delta u'_s / 
\Delta d'_s < 1$ on the non--strange sea distributions, from the expected 
sea spin--component ratio $0 < \Delta(q_Q)_s/\Delta(q_{Q'})_s < 1$ for sea 
partons inside a constituent quark, where the upper and lower bounds are 
justified respectively by the Pauli--principle ``penalty factor'' mentioned 
above and by the similarity in their production mechanisms.

\par With the constraint of Eq. (15) imposed on the polarized valence 
distributions joined with the neglet of the small isovector part in the 
polarized sea ones, the distributions $\delta u_v$ and $\delta d_v$ can be 
normalized to the BjSR, independent of their functional forms, since Eq. (1) 
can be reduced to 
$$
I_0^{p-n}(Q^2) = {1\over6} \cdot \int_0^1 dx ( \delta u_v - \delta d_v ) =
{1\over6} \cdot C_8(\alpha_s) \cdot g_A + (h.t.)_{I_t=1} \ .
\eqno (16)
$$
For these forms, three parametrizations will be adopted, to check the 
systematic effects on the EJSR integrals of the behaviour assumed in the 
isoscalar valence distributions as $x \to 0$.

\par The first parametrization (which will be labeled FRP, for fully 
Reggeized parametrization) breaks down each of the polarized 
distributions for the valence quarks into an isoscalar and isovector 
part, the first dominated at $x \sim 0$ by the $\eta$--meson trajectory, 
and the second by the pion one, so that, after introducing the 
Pauli--principle ``penalty factor'' $(1-x)$ in $\delta d_v$ and 
imposing the conditions in Eqs. (13) and (14), with the $x$--polynomials 
$P_j(x,Q^2)$ again reduced to $Q^2$--dependent coefficients, one gets
$$
\delta u_v^{(1)}(x,Q^2) = C(Q^2) \cdot [ {{231}\over{820}} \cdot 
B({5\over4},4)^{-1} \cdot x^{1\over4} + {{85}\over{41}} ] \cdot (1-x)^3 \ ,
\eqno (17)
$$
$$
\delta d_v^{(1)}(x,Q^2) = C(Q^2) \cdot [ {{231}\over{820}} \cdot 
B({5\over4},4)^{-1} \cdot x^{1\over4} - {{85}\over{41}} ] \cdot (1-x)^4 \ ,
\eqno (17')
$$
where the normalization $C(Q^2) = C_8(\alpha_s) \cdot g_A + 6 \cdot 
(h.t.)_{I=1}$ is 
fixed from Eq. (16) to automatically satisfy the BjSR, and the function 
$B(\alpha,\beta)$ is the well known Euler's beta function.

\par For the second set of model distribution functions (to be labeled SRP, 
or simplified Regge--pole parametrization) one takes the limit 
$\alpha_\eta(0) \to \alpha_\pi(0) \to 0$, still keeping the same constraints, 
so that Eqs. (17), (17$'$) become 
$$
\delta u_v^{(2)}(x,Q^2) = {{16}\over5} \cdot C(Q^2) \cdot (1-x)^3 \ ,
\eqno (18)
$$
$$
\delta d_v^{(2)}(x,Q^2) = - C(Q^2) \cdot (1-x)^4 \ ,
\eqno (18')
$$
where the normalization is again the same. This parametrization 
has the advantage over the previous one of producing an isoscalar 
structure function $g_1^p+g_1^n$ non--vanishing in the limit $x \to 0$, 
but at the cost of violating the isotopic spin properties of the Reggeon 
exchanges.

\par A third set of model distribution functions has been built following 
the suggestion by Bass and Landshoff$^{38}$ that a non--perturbative, 
two--gluon exchange term might develop at small $x$ a behaviour $g_1 \sim 
\log (1/x)$, obviously only in the isoscalar part, for the PDIS 
structure function $g_1$. 
Thus one can write, again with the same number of parameters and the 
same constraints, the functions (labeled, from the names of the 
above--mentioned authors, as BLP) 
$$
\delta u_v^{(3)}(x,Q^2) = C(Q^2) \cdot \bigl( {{66}\over{131}} \log {1\over x} 
+ {{2817}\over{1310}} \bigr) \cdot (1-x)^3 \ ,
\eqno (19)
$$
$$
\delta d_v^{(3)}(x,Q^2) = C(Q^2) \cdot \bigl( {{66}\over{131}} \log {1\over x} 
- {{2817}\over{1310}} \bigr) \cdot (1-x)^4 \ ;
\eqno (19')
$$
this time the $\eta$--meson Regge--pole term has been left out 
to avoid including one 
more parameter and so keep all three parametrizations on equal statistical 
footing. Taking the PQCD coefficients listed above (with the value for 
$\Lambda_{\overline{MS}}(5)$ of Ref. 17) and the HTC of Ref. 9 as inputs, 
there is only one free parameter left for each set of data, {\sl i.e.} 
the sea normalization $P(Q^2)$. Note that, to account for possible 
systematics either in the data or in the theoretical inputs (particularly 
the possible inadequacy both of the PQCD approximation and of the 
phenomenological parametrization used, as well as of the evaluation of 
the HTC), different parameters will be used for different data, even when 
these latter have been normalized at the same value of $<Q^2>$.

\vglue 1.0truecm
\leftline{\bf 4. The isosinglet sum rule and the quark spins in the nucleon. }
\vglue 0.5truecm

\par Before turning to fitting the parametrizations to the data available 
(as of November 1995) on the PDIS structure functions $g_1$, it is better 
to consider the information 
we possess of the axial couplings appearing in the right--hand sides of the 
first--moment sum rules, Eqs. (1) and (4). All pieces of information available 
have been summarized in Table III, which deserves some comments. Usually most 
analyses$^{39}$ of these couplings stop at the information coming from 
asymmetries in the decay products momenta and/or polarizations: while this 
is enough in most of the cases, in some no such measurements are either 
available or possible, and disregard of the information coming from the rates 
can have drastic effects, even on the correctness of the conclusions inferred. 
Enough to say that dropping the information on the $\Delta S = 0$, $\Sigma 
\to \Lambda$ couplings leads to the conclusion that flavour SU(3) 
symmetry breaking effects are negligible$^{39}$, relying in fact {\sl on 
a single point}, being the coupling of the $\Xi$ to both $\Lambda$ and 
$\Sigma$ not as well known as those of the latter two to the nucleon. 
Accordingly, Table III includes evidence from both asymmetries and rates, 
while the 
detailed analysis of these latter can be found elsewhere$^{40}$. One further 
thing evident from Table III is that the two sources of information yield 
fully compatible values for the axial couplings, contrary to the statement 
of Jaffe and Manohar$^{41}$, which originated from a completely outdated 
treatment of the decay rate data.

\vglue 0.6truecm
\centerline{\bf Table III }
\centerline{\bf Data on octet baryon axial couplings }
\vglue 0.3truecm
\hrule
$$\vbox{\halign{#\hfil&\quad\hfil#\hfil&\quad\hfil#\hfil&\quad\hfil#\hfil\cr
Decays & from rates$^{40}$ & from asymmetries$^{42}$ & SU(3) formul\ae$^{43}$ 
\cr
$n \to p e^- \overline{\nu}_e$ & 1.2553 $\pm$ 0.0018 & 1.2573 $\pm$ 0.0028 & 
$F + D$ \cr
$ \Sigma^- \to \Lambda e^- \overline{\nu}_e$ & 0.723 $\pm$ 0.017 & & $D - 
\sqrt3\tan\phi F$ \cr
$ \Sigma^+ \to \Lambda e^+ \nu_e$ & 0.750 $\pm$ 0.094 & & $D + \sqrt3\tan\phi 
F$ \cr
$ \Sigma^- \to n e^- \overline{\nu}_e$ & $-$ 0.330 $\pm$ 0.023 & $-$ 0.340 
$\pm$ 0.017 & $F - D$ \cr
$ \Sigma^- \to n \mu^- \overline{\nu}_\mu$ & $-$ 0.237 $\pm$ 0.078 & & $F - D$ 
\cr
$ \Lambda \to p e^- \overline{\nu}_e$ & 0.729 $\pm$ 0.011 & 0.718 $\pm$ 0.015 
& $F + {1\over3} (1+{4\over{\sqrt3}}\tan\phi) D$ \cr
$ \Lambda \to p \mu^- \overline{\nu}_\mu$ & 0.756 $\pm$ 0.139 & & $F + 
{1\over3} (1+{4\over{\sqrt3}}\tan\phi) D$ \cr
$\Xi^- \to \Lambda e^- \overline{\nu}_e$ & 0.265 $\pm$ 0.044 & 0.250 $\pm$ 
0.050 & $F - {1\over3} (1+{4\over{\sqrt3}}\tan\phi) D$ \cr
$\Xi^- \to \Lambda \mu^- \overline{\nu}_\mu$ & 0.76 $^{+\ 0.47}_{-\ 0.76}$ & & 
$F - {1\over3} (1+{4\over{\sqrt3}}\tan\phi) D$ \cr
$\Xi^- \to \Sigma^o e^- \overline{\nu}_e$ & 1.216 $\pm$ 0.147 & & $F + 
(1-{4\over{\sqrt3}}\tan\phi) D$ \cr}}$$
\hrule
\vglue 0.6truecm

\par Table III does not contain, for reasons of internal clarity, an 
additional piece of information, coming from the study of the asymmetries 
in the $\Sigma^- \to \Lambda$ $\beta$--decay, {\sl i.e.} the ratio$^{42}$ 
$g_V/g_A =$ $-$ 0.01 $\pm$ 0.10, which can be considered a bound on the 
size of the $\Sigma^0 - \Lambda$ mixing effects, which have been included 
following the analysis by Karl$^{43}$, and fixing $\tan\phi$ to the value 
given by him.

\par The SU(3)--symmetric fit to all data listed in Table III, plus 
the ratio $g_V/g_A$ for the decay $\Sigma^- \to \Lambda e^- 
\overline{\nu}_e$, yields the parameters $F =$ 0.4678 and $D =$ 0.7876, 
or $g_8 = 3F - D =$ 0.616 $\pm$ 0.022 and $g_A = F + D =$ 1.2554 $\pm$ 
0.0020 (note that this latter acts almost as a constraint due to the very 
high precision of neutron data on both asymmetries and rates); the $\chi^2$ 
of the fit is not very high with respect to the number of data points, 
being 20.17 (about 2 units better than {\sl without} $\Sigma^0$--$\Lambda$ 
mixing) versus 15, but is concentrated almost exclusively in the $\Sigma^- 
\to \Lambda$ rate, whose datum on the axial coupling lies $3\sigma$ 
below the SU(3)--symmetric fit value. Coming this datum from a 
good--quality experiment$^{44}$, this fact has to be taken as 
initial evidence for some flavour SU(3) breaking in these couplings, being 
the mixing required to explain this discrepancy more than five times 
the one calculated by Karl$^{43}$, and thus giving a ratio $g_V/g_A$ more 
than twice its experimental $1\sigma$ limit. This must be remembered 
when using flavour symmetry to extract the nucleon spin composition. 
A further point 
raised$^{30}$ in this context is also the possible effect (particularly on 
the $\Delta S = 1$ transitions) of the tensor components in the weak axial 
current: the good agreement between the two columns of data in Table III 
seems to indicate that the effect is not as large as indicated by some 
fits$^{45}$, since it would affect the axial couplings extracted from the 
asimmetries differently from those extracted from the rates.

\par Since it is the aim of the present paper to concentrate on the analysis 
of the spin composition of the nucleon, the PDIS structure functions 
will not be fit 
treating both $C(Q^2)$ and $P(Q^2)$ as free parameters, but rather taking the 
first from the right--hand side of the BjSR, Eq. (1), and fitting only the 
second to the data, experiment by experiment. Table IV will present 
the expectations for the right--hand side of the BjSR in NNLO PQCD, 
including the HTC of Ross and Roberts$^9$, together with the integrals 
evaluated by two 
experimental groups, the E143 Collaboration at SLAC$^{13}$ and the Spin Muon 
Collaboration at CERN$^{46}$. In the same table, we shall also display the two 
expectations for the right--hand side of the EJSR obtained assuming 
validity of the OZI rule, {\sl i.e.} $\Sigma = g_8$, and either a) 
the na\"{\i}ve parton model identification $g_0 = \Sigma$ ({\sl i.e.} 
decoupling the contribution from the axial anomaly, as is the case 
for {\sl massive} quarks), or b) the definition in Eq. (5) with the 
normalizations scale $\mu^2$ fixed so that $\alpha_s(\mu^2) = 1$ 
(and $<L_z>=0$): here 
$g_8$ is assumed equal to the above flavour SU(3) symmetry prediction. 
The choice a) amounts to using an isoscalar, PQCD corrected version of the 
{\sl original} EJSR$^{20}$, while b) serves to give an idea of the 
effect induced by a very reasonably sized axial anomaly 
contribution$^{21,22}$.

\par From this table one can read two facts, already mentioned above: the 
running with $Q^2$ of the EJSR is slower than that of the BjSR, and 
therefore harder to see in the data unless higher precisions are 
reached, and the difference between the ``anomalous'' and ``non--anomalous'' 
versions of the EJSR, described by Eq. (8) (apart from the lower 
asymptotic value of the first, which can always be traded for a 
breaking of the OZI rule, {\sl i.e.} a $\Delta s \neq 0$, in the 
second), is even smaller, and thus much harder to see.

\vfill
\eject
\vglue 0.6truecm
\centerline{\bf Table IV }
\centerline{\bf Sum rule expectations for BjSR and na\"{\i}ve EJSR }
\vglue 0.3truecm
\hrule
$$\vbox{\halign{\hfil # \hfil & \quad \hfil # \hfil & \quad \hfil # \hfil & 
\quad \hfil # \hfil & \quad \hfil # \hfil & \quad # \hfil \cr
$Q^2$/GeV$^2$ & EJSR (a) & EJSR (b) & BjSR & BjSR integrals & Ref. \cr
2.0 & 0.1278 & 0.0917 & 0.1492 &  &  \cr
3.0 & 0.1378 & 0.0993 & 0.1636 & 0.163 $\pm$ 0.019 & 13 \cr
4.6 & 0.1448 & 0.1045 & 0.1733 &  &  \cr
6.0 & 0.1479 & 0.1068 & 0.1776 &  &  \cr
8.0 & 0.1506 & 0.1088 & 0.1813 &  &  \cr
10.0 & 0.1523 & 0.1099 & 0.1836 & 0.194 $\pm$ 0.038 & 46 \cr
10.7 & 0.1528 & 0.1103 & 0.1843 &  &  \cr
12.0 & 0.1535 & 0.1108 & 0.1853 &  &  \cr
}}$$
\hrule
\vglue 0.6truecm

\par In Table V there will be displayed the parameters $P(Q^2)$ 
obtained from the fits to the seven sets of PDIS data analyzed here: the 
re--normalized EMC set of data on proton$^{1,5}$ at $<Q^2> =$ 10.7~GeV$^2$, 
the E142 neutron data$^4$ at $<Q^2> =$ 2.0~GeV$^2$, the SMC preliminary 
deuteron data$^{14}$ at $<Q^2>=$ 4.6~GeV$^2$, the SMC proton data$^{24}$ 
at $<Q^2>=$ 10.0~GeV$^2$, the E143 data on proton$^{12}$ and deuteron$^{13}$ 
at $<Q^2>=$ 3.0~GeV$^2$, and the SMC new deuteron data$^{40}$ at $<Q^2>=$ 
10.0~GeV$^2$. The very preliminary data of the SLAC--Yale E80 and E130 
Collaborations$^{47}$ have not been included in the fits, since they cover 
either very low values of $Q^2$ or a very limited range in $x$ values, 
and would have been, besides, of very little statistical significance; 
also, the recently appeared E143 data$^{28}$ are not included, since 
they have become available only during the final redaction of 
the present paper. The factor $1/2 \cdot(1-{3\over2}\omega_D)$, with 
$\omega_D=$ 0.05, has been explicitly included for deuteron data, so that
all $P(Q^2)$ obtained from the fits have the same normalization. All are 
relatively large and negative, 
implying a strong reduction from the pure valence contribution to the EJSR, 
which under our constraint Eq. (15) would be $I_0^{p+n}(Q^2)_{val} = 
I_0^{p-n}(Q^2)_{val} = {1\over6} \cdot C(Q^2)$, {\sl i.e.} equal to the BjSR 
integral: from this and the previous table one can see that some negative 
sea is already expected even 
at the level of a na\"{\i}ve formulation of the 
(PQCD corrected) EJSR. The table displays also systematic variations in the 
parameter both with the model used for the valence distributions and with the 
nature of the target, variations which in the opinion of the author cast some 
doubt on the possibility of really testing PQCD in the isoscalar combination: 
since the two kind of variations are of the same order of magnitude, 
and comparable to the statistical errors from the fits, their origin 
is difficult to trace.

\vfill
\eject
\vglue 0.6truecm
\centerline{\bf Table V }
\centerline{\bf Values of $P(Q^2)$ from fits to the data }
\centerline{ (shown in parenthesis are the $\chi^2$--values per data point) }
\vglue 0.3truecm
\hrule
$$\vbox{\halign{# \hfil & \hfil # \hfil & \hfil # \hfil & \hfil # \hfil & 
\hfil # \hfil \cr
Expt. & $Q^2$/GeV$^2$ & FRP fit & SRP fit & BLP fit \cr
E142$^n$ & 2.0 & $-0.38 \pm 0.07$ (0.7) & $-0.35 \pm 0.07$ (0.5) & $-0.23 \pm 
0.13$ (3.1) \cr
E143$^p$ & 3.0 & $-0.69 \pm 0.04$ (1.9) & $-0.72 \pm 0.05$ (2.2) & $-0.73 \pm 
0.08$ (6.2) \cr
E143$^d$ & 3.0 & $-0.53 \pm 0.04$ (1.3) & $-0.54 \pm 0.04$ (1.2) & $-0.51 \pm 
0.08$ (4.7) \cr
SMC$^d$ & 4.6 & $-0.53 \pm 0.24$ (0.6) & $-0.59 \pm 0.24$ (0.7) & $-0.66 \pm 
0.27$ (1.1) \cr
SMC$^p$ & 10.0 & $-0.70 \pm 0.18$ (1.7) & $-0.75 \pm 0.14$ (1.0) & $-0.80 \pm 
0.13$ (0.7) \cr
SMC$^d$ & 10.0 & $-0.44 \pm 0.18$ (2.7) & $-0.49 \pm 0.18$ (2.5) & $-0.53 \pm 
0.22$ (3.8) \cr
EMC$^p_{rev}$ & 10.7 & $-0.74 \pm 0.18$ (0.4) & $-0.79 \pm 0.18$ (0.8) & 
$-0.82 \pm 0.20$ (1.1) \cr
$\chi^2/N_{pts}$ & & 149.7 / 102 & 152.3 / 102 & 373.7 / 102 \cr
}}$$
\hrule
\vglue 0.6truecm

\par These results for the sea parameter $P(Q^2)$ can now be turned 
into values for the EJSR integrals, since these can be written as 
$$
I_0^{p+n}(Q^2) = {1\over6} \cdot C(Q^2) + 2 \cdot B({5\over4},8) 
\cdot P(Q^2) \ .
\eqno (20)
$$
Note however that the right--hand side of this equation corresponds to what is 
actually interpolated by the fits only in the case of a deuterium target, when 
$I_0^d(Q^2) = {1\over2} \cdot (1-{3\over2}\omega_D) \cdot I_0^{p+n}(Q^2)$: for 
a proton or neutron target the fits measure, respectively, 
$$
I_0^p(Q^2) = {1\over6} \cdot C(Q^2) + B({5\over4},8) \cdot P(Q^2)
\eqno (20')
$$
in the proton case, and 
$$
I_0^n(Q^2) = B({5\over4},8) \cdot P(Q^2)
\eqno (20'')
$$
in the neutron one, to which one must, respectively, either subtract or add 
the BjSR right--hand side: 
$$
I_0^{p+n}(Q^2) = 2\cdot I_0^p(Q^2) - {1\over6} \cdot C(Q^2)
\eqno (21)
$$
in the first case, and 
$$
I_0^{p+n}(Q^2) = 2\cdot I_0^n(Q^2) + {1\over6}\cdot C(Q^2)
\eqno (21')
$$
in the second. In these cases the last tems are no longer part of the valence 
distribution normalization, and therefore they carry a theoretical error, 
difficult to quantify$^{48}$, but approximable with that of the most precise 
experimental determination of the BjSR integral ({\sl i.e.} with that of 
Ref. 46). 

\par Taking these aspects into account, the fits yield for the EJSR the 
integrals listed in Table VI for the three valence parametrizations, which 
show a marked reduction with respect the na\"{\i}ve expectations for this 
quantity, tabulated in the second column of Table IV. As already said, 
this is evidence for either presence of a sizeable reduction of $g_8$ 
and/or $g_0$ (due to a polarized strange--quark density or the breaking 
of flavour SU(3) symmetry$^{49}$) or an appreciable contribution from 
the axial anomaly (or both).

\par From the previous table one can also see that the three parametrizations, 
though behaving quite differently at $x \to 0$ in the combination $g_1^{p+n}$, 
do not produce neither appreciably different $\chi^2$'s (the BLP fit turns 
worse than the other two, wich statistically are on the same level, but 
only by a factor $\sim$ 2.5), nor large variations (contrary to na\"{\i}ve 
expectations) for $I_0^{p+n}(Q^2)$: this is due to the normalization of 
the valence components to the BjSR imposed by Eq. (15)--(16), valid for 
all parametrizations regardless of their behaviour as $x \to 0$ and quite 
robust, since the only isovector contribution from the sea could come, under 
the hypotheses adopted here, from the Pauli principle ``penalty factor''. 
The different behaviours as $x \to 0$ of the valence parametrizations 
reflect thus only on the quality of the fits, tending to prefer distributions 
which stay finite in this limit, since diverging ones become harder to 
accommodate to the behaviours of $g_1$ for neutron and deuteron targets 
at moderate and high values of the Bjorken variable $x$.

\vglue 0.6truecm
\centerline{\bf Table VI }
\centerline{\bf Integrals of the EJSR from fits to the data }
\vglue 0.3truecm
\hrule
$$\vbox{\halign{# \hfil & \hfil # \hfil & \quad \hfil # \hfil & \quad \hfil # 
\hfil & \quad \hfil # \hfil \cr
Expt. & $Q^2$/GeV$^2$ & FRP fit & SRP fit & BLP fit \cr
E142$^n$ & 2.0 & 0.098 $\pm$ 0.020 & 0.103 $\pm$ 0.020 & 0.118 $\pm$ 0.025 \cr
E143$^p$ & 3.0 & 0.073 $\pm$ 0.020 & 0.068 $\pm$ 0.020 & 0.068 $\pm$ 0.022 \cr
E143$^d$ & 3.0 & 0.094 $\pm$ 0.005 & 0.092 $\pm$ 0.005 & 0.096 $\pm$ 0.010 \cr
SMC$^d$ & 4.6 & 0.103 $\pm$ 0.032 & 0.095 $\pm$ 0.032 & 0.086 $\pm$ 0.036 \cr
SMC$^p$ & 10.0 & 0.090 $\pm$ 0.032 & 0.084 $\pm$ 0.028 & 0.077 $\pm$ 0.028 \cr
SMC$^d$ & 10.0 & 0.124 $\pm$ 0.024 & 0.119 $\pm$ 0.023 & 0.114 $\pm$ 0.029 \cr
EMC$^p_{rev}$ & 10.7 & 0.086 $\pm$ 0.032 & 0.080 $\pm$ 0.032 & 0.075 $\pm$ 
0.034 \cr
}}$$
\hrule
\vglue 0.6truecm

\par Any interpretation of these data (beyond the existence of a sizeable, 
negative polarization in the sea) has to make clear the nature of the 
unitary--singlet piece in the EJSR. If one follows the na\"{\i}ve parton 
model picture to the end, and identifies $g_0$ with $\Sigma$, one must 
also have $\tilde \gamma =$ 0 from the vanishing of its anomalous 
dimension, so that the EJSR becomes now, since $g_8 = \Sigma - 3 \Delta s$, 
$$
I_0^{p+n}(Q^2) = [{2\over9} \cdot C_1(\alpha_s) + {1\over{18}} \cdot 
C_8(\alpha_s) ] \cdot \Sigma - {1\over6} \cdot C_8(\alpha_s) \cdot \Delta s + 
(h.t.)_{I_t=0} \ ,
\eqno (22)
$$
which, putting $\delta I = - (h.t.)_{I_t=0}$, $F(Q^2) = {2\over9} \cdot 
C_1(\alpha_s) + {1\over{18}} \cdot C_8(\alpha_s)$ and $R(Q^2) = C_8(\alpha_s)/
[6 \cdot F(Q^2)]$, can be turned into a linear relation between $\Sigma$ and 
$\Delta s$,
$$
\Sigma = {{I_0^{p+n}(Q^2) + \delta I}\over{F(Q^2)}} + R(Q^2) \cdot \Delta s 
= \Sigma_0(Q^2) + R(Q^2) \cdot \Delta s \ ,
\eqno (23)
$$
to be eventually combined with the flavour SU(3) {\sl prediction} for $g_8 = 
\Sigma - 3 \Delta s$ (and with $g_A = \Delta u - \Delta d$) to disentangle 
the quark spin components in the proton ({\sl modulo} flavour SU(3) symmetry 
violation effects).

\par Though simple and attractive, Eq. (22) runs however against 
the PQCD prediction on the evolution of {\sl light} parton 
densities$^{22}$: when the mass--parameter of these latter 
is neglected, the gluon--to--quark splitting function dictates 
for $g_0$ an anomalous evolution and at the same time identifies it with 
the combination in Eq. (5). One is now faced with the additional problem 
of choosing a normalization scale $\mu^2$ to relate it to the spin 
composition of the nucleon. For this, let us consider the angular 
momentum sum rule
$$
\Sigma + 2 \cdot [ \Delta G(Q^2) + <L_z> ] = 1 \ ,
\eqno (24)
$$
which has to be satisfied at all momentum scales. Here both $\Delta G$ and 
$<L_z>$ must run with the momentum scale, and from the evolution equation for 
$g_0$ one can see that the first will run like
$$
\Delta G(Q^2) = {{\alpha_s(\mu^2)}\over{\alpha_s(Q^2)}} \cdot \exp [ \tilde 
\gamma (\alpha_s(Q^2)) - \tilde\gamma(\alpha_s(\mu^2)) ] \cdot \bigl\{ \Delta 
G(\mu^2) + 
$$
$$
+ {{2\pi}\over{3\alpha_s(\mu^2)}} \cdot \Sigma \cdot \bigl( \exp [ 
\tilde \gamma (\alpha_s(\mu^2)) - \tilde\gamma(\alpha_s(Q^2)) ] - 1\bigr) 
\bigr\} \ ,
\eqno (25)
$$
while $<L_z> = (1-\Sigma)/2-\Delta G(Q^2)$, where the first term 
$(1-\Sigma)/2$ is positive in all reasonably conceivable models of the 
nucleon spin structure.

\par It is therefore ``natural'' to assume that $<L_z>$ will vanish at some 
low mass scale, where $\Sigma < 1$ will be balanced by some polarized glue 
distributed between the constituent quarks. If we take $\mu^2$ as such a 
scale, then $\Delta G(\mu^2)=(1-\Sigma)/2$ and the EJSR becomes
$$
I_0^{p+n}(Q^2) = \bigl\{ {2\over9} \cdot \bigl[ 1 + 
{{3\alpha_s(\mu^2)}\over{4\pi}} \bigr] \cdot \exp[\tilde\gamma(\alpha_s) - 
\tilde\gamma(\alpha_s(\mu^2))] \cdot C_1(\alpha_s) + {1\over{18}} \cdot 
C_8(\alpha_s) \bigr\} \cdot \Sigma - 
$$
$$
- {1\over6} \cdot C_8(\alpha_s) \cdot \Delta s + (h.t.)_{I_t=0} - 
{{\alpha_s(\mu^2)}\over{6\pi}} \cdot \exp [ \tilde 
\gamma(\alpha_s) - \tilde \gamma(\alpha_s(\mu^2)) ] \cdot C_1(\alpha_s) \ 
\eqno (26)
$$

\par The linear relation between $\Sigma$ and $\Delta s$, Eq. (23) is still 
holding, but now $F(Q^2)$ is changed including the integrated anomalous 
dimension factor $[1+3\alpha_s(\mu^2)/4\pi]\cdot\exp[\tilde\gamma(\alpha_s) - 
\tilde\gamma(\alpha_s(\mu^2))]$ to multiply the PQCD coefficient 
$C_1(\alpha_s)$, and to the correction $\delta I$ one has to add a {\sl 
positive} contribution --- equal to minus the last term in Eq. (26) --- 
coming again from the anomaly contribution at the normalization scale 
$\mu^2$. This last term becomes thus comparable to the EJSR integral 
$I_0^{p+n}(Q^2)$, so that it is not too difficult in this approach to 
recover a quark content of the nucleon spin very close to the flavour SU(3) 
prediction for $g_8$, as people were na\"{\i}vely expecting before the EMC 
results$^{1,5}$ became public.

\par Table VII lists the numerical values for the terms $\delta I(Q^2)$, 
$F(Q^2)$ and $R(Q^2)$ in Eq. (23) under the two hypotheses: for the second 
one, $\mu^2$ is selected here by putting $\alpha_s(\mu^2) = 1$, a not 
unreasonable choice in view of the fact that we expect it to lie at 
the bottom end of the PQCD validity range, due to the smallness of the 
$\Delta G$ involved, whose size at a scale $Q^2\simeq$ 1 $\sim$ 2~GeV$^2$ is 
expected$^{34}$ to be comparable with $<x_G>\simeq$ 1/2. One can note that 
the PQCD evolution has its largest effect in the factor $F(Q^2)$, 
rescaling $I_0^{p+n}(Q^2)$ to $\Sigma_0$, which in the parton model limit 
should take the value 5/18, while $R(Q^2)$ in both cases considered 
deviates from its parton model value 3/5 only at the lowest values of $Q^2$.

\vglue 0.6truecm
\centerline{\bf Table VII }
\centerline{\bf Numerical values for the terms appearing in Eq. (23) }
\vglue 0.3truecm
\hrule
$$\vbox{\halign{\hfil # \hfil & \quad \hfil # \hfil & \quad \hfil # \hfil & 
\quad \hfil # \hfil \cr
$Q^2$/GeV$^2$ & $F(Q^2)$ & $\delta I$ & $R(Q^2)$ \cr
  & Eq. (22)$\ \ $Eq. (26) & Eq. (22)$\ \ $Eq. (26) & Eq. (22)$\ \ $Eq. (26) 
\cr
2.0 & 0.2310 $\ \ \ $ 0.2315 & 0.0145 $\ \ \ $ 0.0509 & 0.5566 $\ \ \ $ 0.5553 
\cr
3.0 & 0.2394 $\ \ \ $ 0.2368 & 0.0097 $\ \ \ $ 0.0465 & 0.5712 $\ \ \ $ 0.5775 
\cr
4.6 & 0.2453 $\ \ \ $ 0.2403 & 0.0063 $\ \ \ $ 0.0435 & 0.5795 $\ \ \ $ 0.5915 
\cr
10.0 & 0.2520 $\ \ \ $ 0.2442 & 0.0029 $\ \ \ $ 0.0404 & 0.5874 $\ \ \ $ 
0.6062 \cr
10.7 & 0.2525 $\ \ \ $ 0.2444 & 0.0027 $\ \ \ $ 0.0403 & 0.5878 $\ \ \ $ 
0.6072 \cr
}}$$
\hrule
\vglue 0.6truecm

\par Using these values one can extract the terms $\Sigma_0(Q^2)$ in the 
linear relationship of Eq. (23), $\Sigma = \Sigma_0(Q^2) + R(Q^2) \cdot 
\Delta s$, listed in Table VIII. 

\par To sum the results into single values, the different experiments can be 
averaged {\sl \`a la} PDG, so as to weight their different (both experimental 
and theoretical) accuracies: the outcomes are 
$$
\Sigma = (0.418 \pm 0.021) + 0.5582 \cdot \Delta s
\eqno (27)
$$
for the ``non--anomalous'' treatment of the EJSR, while the correct inclusion 
of the axial anomaly (with the choices $\alpha_s=1$, $<L_z>=0$ at its 
normalization scale $Q^2=\mu^2$) leads to an average 
$$
\Sigma = (0.579 \pm 0.022) + 0.5796 \cdot \Delta s \ .
\eqno (28)
$$

\vglue 0.6truecm
\centerline{\bf Table VIII }
\centerline{\bf Values of $\Sigma_0$ from the EJSR data }
\centerline{ (averaged over the three fits) }
\vglue 0.3truecm
\hrule
\vglue 0.3truecm
$$\vbox{\halign{# \hfil & \hfil # \hfil & \quad \hfil # \hfil & \quad \hfil # 
\hfil \cr
expt. & $Q^2$/Gev$^2$ & From Eq. (22) & From Eq. (26) \cr
E142$^n$ & 2.0 & 0.518 $\pm$ 0.091 & 0.674 $\pm$ 0.091 \cr
E143$^p$ & 3.0 & 0.329 $\pm$ 0.060 & 0.488 $\pm$ 0.061 \cr
E143$^d$ & 3.0 & 0.429 $\pm$ 0.025 & 0.590 $\pm$ 0.026 \cr
SMC$^d$ & 4.6 & 0.415 $\pm$ 0.134 & 0.578 $\pm$ 0.137 \cr
SMC$^p$ & 10.0 & 0.341 $\pm$ 0.115 & 0.506 $\pm$ 0.118 \cr
SMC$^d$ & 10.0 & 0.487 $\pm$ 0.100 & 0.656 $\pm$ 0.103 \cr
EMC$^p_{rev}$ & 10.7 & 0.331 $\pm$ 0.130 & 0.495 $\pm$ 0.134 \cr
}}$$
\hrule
\vglue 0.6truecm

\par One can note that the treatments proposed here lead {\sl in both cases} 
to higher values for $\Sigma$ than those to be found in the recent 
literature$^{6,15,27,50}$, though of course in the second case much 
higher than in the first. It also noteworthy (in the author's opinion) 
that the second average for $\Sigma_0$ is remarkably close 
to the flavour SU(3) value derived above for $g_8$, since estimates on its 
flavour--symmetry breaking part led to expect a value slightly lower than 
the symmetry prediction$^{49}$.

\par Note also that this averaging does not produce results much different 
from those obtainable selecting the deuterium data alone, which should be 
better from the point of view of theoretical systematics, since in the 
handling of the data one does not have to correct them with the BjSR 
integrals. The reduction in the scale of the error is of course a product 
of the use of the full statistics of the (almost) complete world data set.

\par Intersecting the above error bands on $\Sigma$ and $\Delta s$ with 
that from the flavour SU(3) result 
$$
g_8 = \Sigma - 3 \cdot \Delta s = 0.616 \pm 0.022 \ ,
\eqno (29)
$$
one can now obtain the results ({\sl modulo} of course SU(3) violations) 
$$
\Sigma = 0.373 \pm 0.026 \ \ \ {\rm and} \ \ \ \Delta s = -0.081 \pm 0.012 
\eqno (30)
$$
for the ``non--anomalous'' treatment of the EJSR, and 
$$
\Sigma = 0.570 \pm 0.028 \ \ \ {\rm and} \ \ \ \Delta s = -0.015 \pm 0.013 \ ,
\eqno (31)
$$ 
for the description of $g_0(Q^2)$ including the axial anomaly, 
not very far from the na\"{\i}ve OZI expectations $\Sigma \simeq g_8$ and 
$\Delta s \simeq 0$, and practically consistent with them, if the theoretical 
uncertainties$^{49}$ entailed by the use of flavour SU(3) symmetry are duly 
considered.

\par Comparing the present analysis with the recently published 
ones$^{15,27,50}$, the values of Eq. (31) reproduce within 1$\sigma$ those 
for the asymptotic limit of the unitary--singlet axial charge 
$g_0(\infty)^{N_f=3}$, misnamed $\Sigma$ in much of the current 
literature, 
$$
g_0(\infty)^{N_f=3} = \exp - \tilde\gamma(\alpha_s(\mu^2)) \cdot [ \Sigma - 
{{3\alpha_s(\mu^2)}\over{4\pi}} (1 - \Sigma) ] = 0.342 \pm 0.025 \ ,
\eqno (32)
$$
but note also that neglecting the anomaly (and trading it 
for a sizeable $\Delta s$) we find a value for the spin sum $\Sigma$ 
still 20\% higher than the currently quoted (and, in the author's 
opinion, erroneously derived) values. Note, as well, that, unless one 
assumes SU(3) to be a good symmetry for the axial charges, one can 
not conclude that the low $\Sigma_0$ obtained with the identification 
$g_0 = \Sigma$ is evidence for a large and negative $\Delta s$, since 
the value 0.418 $\pm$ 0.021 lies inside the allowed 
range of flavour--SU(3)--violating effects expected$^{49}$ for $g_8$.

\par A test of these two possible ways out of the conundrum on the 
smallness of $g_0$, in the range of $Q^2$ explored by the PDIS 
experiments, is offered by the possibility of measuring the size and 
sign of $\Delta G(Q^2)$, on which the ``anomalous'' interpretation 
gives sharp predictions: indeed with the above normalizations one can 
rewrite Eq. (25) as
$$
\Delta G(Q^2) = {{\alpha_s(\mu^2)}\over{\alpha_s(Q^2)}} \cdot \exp [ \tilde 
\gamma(\alpha_s(Q^2)) - \tilde\gamma(\alpha_s(\mu^2))] \cdot 
$$
$$
\cdot \bigl\{ {1\over2} 
+ \Sigma \cdot \bigl[ {{2\pi}\over{3\alpha_s(\mu^2)}} \cdot ( \exp [ \tilde 
\gamma(\alpha_s(\mu^2)) - \tilde\gamma(\alpha_s(Q^2))] - 1) - {1\over2} \bigr] 
\bigr\}
\eqno (33)
$$
which, due to the smallness of the coefficient of $\Sigma$ for $Q^2 \ge$ 
2~GeV$^2$, gives quite accurate predictions on the gluonic polarization 
$\Delta G$ in the nucleon, tabulated below for the value of $\Sigma$ 
given by Eq. (31): these predictions could be tested, for instance, 
measuring the polarization asymmetries for the semi--inclusive process 
$\ell^\pm N \to \ell^\pm J/\psi + X$, probably a much clearer signature 
for $\Delta G$ than that proposed via semi--inclusive kaon deep--inelastic 
production$^{30}$ to test the size and sign of $\Delta s$.

\par The third column in this last table displays clearly the effect 
of the higher orders of PQCD in lowering the values of $\Delta G$ at 
low $Q^2$ with respect to the leading--order prediction, which would 
lead to a constant value, even higher than the asymptotic limit listed 
in the last line, due to the large higher--order terms in the 
coefficients $C_{1,8}(\alpha_s)$ (see Table II).

\vfill
\eject
\vglue 0.6truecm
\centerline{\bf Table IX }
\centerline{\bf Values of $\Delta G$ and $\alpha_s \Delta G$ from the anomaly }
\centerline{\bf contribution to the EJSR and $\Sigma$ from Eq. (31) }
\vglue 0.3truecm
\hrule
\vglue 0.3truecm
$$\vbox{\halign{\hfil # \hfil & & \quad \hfil # \hfil & \quad \hfil # \hfil 
\cr
$Q^2$/Gev$^2$ & $\Delta G$ & $\alpha_s \Delta G$ \cr
$\mu^2$ & 0.215 $\pm$ 0.014 & 0.215 $\pm$ 0.014 \cr
1.5 & 0.8453 $\pm$ 0.0026 & 0.3877 $\pm$ 0.0012 \cr
2.0 & 0.9996 $\pm$ 0.0004 & 0.4014 $\pm$ 0.0002 \cr
3.0 & 1.2025 $\pm$ 0.0023 & 0.4142 $\pm$ 0.0008 \cr
4.6 & 1.4030 $\pm$ 0.0049 & 0.4234 $\pm$ 0.0015 \cr
6.0 & 1.5224 $\pm$ 0.0064 & 0.4277 $\pm$ 0.0018 \cr
8.0 & 1.6843 $\pm$ 0.0079 & 0.4316 $\pm$ 0.0021 \cr
10.0 & 1.7439 $\pm$ 0.0091 & 0.4342 $\pm$ 0.0023 \cr
10.7 & 1.7725 $\pm$ 0.0095 & 0.4349 $\pm$ 0.0023 \cr
12.0 & 1.8208 $\pm$ 0.0101 & 0.4361 $\pm$ 0.0024 \cr
$\infty\ (N_f=3)$ & $\infty$ & 0.4778 $\pm$ 0.0055 \cr
}}$$
\hrule
\vglue 0.6truecm

\vglue 1.0truecm
\leftline{\bf 4. Summary and conclusions. }
\vglue 0.5truecm

\par The several experiments now conducted on the PDIS asymmetries (SLAC 
experiments E142$^4$ and E143$^{12,13}$, and the collaborations EMC$^{1,5}$ 
and SMC$^{14,24,46}$ at CERN) do not contradict conventional expectations 
on the spin structure of the nucleon, namely that one should have a 
strange spin component $\Delta s$ much smaller than the non--strange--quark 
sea ones on one side, but on the other no pure valence--quark spin 
components either, as known since more than twenty years (and a complete 
list of references would be almost as long as this paper: it is enough 
to refer to the recent, illuminating papers by Lipkin$^{51}$) from the 
reduction in $g_A$ with respect to the na\"{\i}ve constituent quark model 
prediction $g_A = 5/3$.

\par One finds that it is possible to reduce drastically $g_0(Q^2)$ from 
the parton model plus OZI rule expectation $g_0 \simeq g_8$, due {\sl both} 
to the presence of the QCD axial anomaly$^{22}$ {\sl and} to its anomalous 
dimension$^{23}$ (and more to the second than to the first reason): a {\sl 
correct} use of QCD with high orders included (which make these effects 
even larger than the next--to--leading--order alone) is necessary to 
describe the EJSR without conflict with both the experiments and our 
expectations. The ``spin crisis'' of 1988 was the result as much of an 
inadequate theoretical description as of the low normalization in the 
$F_2^p$ values used by the EMC.

\par However, the presence of the anomaly is not {\sl strictly required} 
by the data$^{52}$: from a purely statistical point of view, these could 
be described also by trading it for a sizeable (not necessarily strange) 
sea component in the nucleon spin: but an adequate parton model description 
excluding the contribution of the anomaly has also to exclude the 
anomalous dimension it carries (for the diagonalization of the evolution 
matrix for the pair $\Sigma$ and $\Delta G$ requires one of the eigenvalues 
to vanish), thus increasing by as much as 20\% the total spin sum $\Sigma$ 
over currently quoted values, and making it possible to explain the data 
through a sizeable, though not impossible, reduction in $g_8 \simeq g_0 = 
\Sigma$ due to flavour--SU(3)--breaking effects$^{49}$.

\par Direct measurements of either a large $\Delta G$ or a large $\Delta s$ 
--- or even of the absence of both, in the case of a large, negative 
SU(3)--breaking effect --- are clearly called for to make a definitive 
choice between these possible ways out of the ``nucleon spin crisis''.

\vglue 1.0truecm
\leftline{\bf 5. Acknowledgements. }
\vglue 0.5truecm

\par The author is grateful to many colleagues for discussions on the points 
touched in the present paper, and in particular to his Perugia colleagues and 
dear friends C. Ciofi degli Atti, A. Masiero and Y.N. Srivastava for their 
encouraging support.

\vglue 1.0truecm
\centerline{\bf REFERENCES AND FOOTNOTES }
\vglue 0.5truecm

\item{1.} {European Muon Collaboration (J.~Ashman, {\sl et al.}): {\sl Phys. 
Lett.} {\bf B 206} (1988) 364.}

\item{2.} {J.D.~Bjorken: {\sl Phys. Rev.} {\bf 148} (1966) 1467; {\sl Phys. 
Rev.} {\bf D 1} (1970) 1376.}

\item{3.} {Ya.Ya.~Balitski\u\i, V.M.~Braun and A.V.~Kolesnichenko: {\sl JETP 
Lett.} {\bf 50} (1989) 61; {\sl Phys. Lett.} {\bf B 242} (1990) 245, {\sl 
erratum} {\bf B 318} (1993) 648. See also Refs. 9--10 for more recent 
re--evaluations.}

\item{4.} {SLAC E142 Collaboration (P.L.~Anthony, {\sl et al.}): {\sl Phys. 
Rev. Lett.} {\bf 71} (1993) 959.}

\item{5.} {European Muon Collaboration (J.~Ashman, {\sl et al.}): {\sl Nucl. 
Phys.} {\bf B 328} (1989) 1.}

\item{6.} {J.~Ellis and M.~Karliner: {\sl Phys. Lett.} {\bf B 313} (1993) 131; 
{\sl ``PAN XIII. Particles and Nuclei, Proc. of the 13th Int. Conf., Perugia 
1993''}, ed. by A.~Pascolini (World Scientific, Singapore 1994), p. 48.}

\item{7.} {R.L.~Workman and R.A.~Arndt: {\sl Phys. Rev.} {\bf D 45} (1992) 
1789; V.D.~Burkert and B.L.~Ioffe: {\sl Phys. Lett.} {\bf B 296} (1992) 223; 
V.D.~Burkert and Z.--J.~Li: {\sl Phys. Rev.} {\bf D 47} (1993) 46; V.~Bernard, 
N.~Kaiser and U.--G.~Mei{\ss}ner: {\sl Phys. Rev.} {\bf D 48} (1993) 3062.}

\item{8.} {S.B.~Gerasimov: {\sl Yad. Fiz.} {\bf 2} (1965) 598; {\bf 5} (1967) 
1263; S.D.~Drell and A.C.~Hearn: {\sl Phys. Rev. Lett.} {\bf 16} (1966) 908.}

\item{9.} {G.G.~Ross and R.G.~Roberts: {\sl Phys. Lett.} {\bf B 322} (1994) 
425.}

\item{10.} {E.~Stein, P.~Gornicki, L.~Mankiewicz and A.~Sch\"afer: {\sl Phys. 
Lett.} {\bf B 343} (1995) 369; {\bf B 353} (1995) 107.}

\item{11.} {S.A.~Larin and J.A.M.~Vermaseren: {\sl Phys. Lett.} {\bf B 259} 
(1991) 345.}

\item{12.} {SLAC E143 Collaboration (K.~Abe, {\sl et al.}): {\sl Phys. Rev. 
Lett.} {\bf 74} (1995) 346.}

\item{13.} {SLAC E143 Collaboration (K.~Abe, {\sl et al.}): {\sl Phys. Rev. 
Lett.} {\bf 75} (1995) 25.}

\item{14.} {Spin Muon Collaboration (B.~Adeva, {\sl et al.}): {\sl Phys. 
Lett.} {\bf B 302} (1993) 533.}

\item{15.} {See the BjSR analysis of J.~Ellis and M.~Karliner: {\sl Phys. 
Lett.} {\bf B 341} (1995) 397.}

\item{16.} {A.L.~Kataev: {\sl Phys. Rev.} {\bf D 50} (1994) R 5469. This are 
the PQCD corrections also employed by Ref. 15, although in a way slightly 
different from here.}

\item{17.} {S.~Bethke and S.~Catani: preprint CERN--TH. 6484/92 (Geneve 1992), 
summary talk presented at the ``27th Rencontre de Moriond'', Les Arcs, March 
1992.}

\item{18.} {Besides PDIS data --- see the analysis in Ref. 15 --- the only 
real improvements on $\alpha_s$ come from the analyses of the HERA data: 
these however confirm essentially the results of Ref. 17. See, for 
instance, the report from the ZEUS Collaboration (M.~Derrick, {\sl et 
al.}): preprint DESY 95--182 (Hamburg, October 1995), also available as 
hep--ex/9510001.}

\item{19.} {See, for instance: G.~Preparata and P.G.~Ratcliffe: {\sl EMC, 
E142, SMC, Bjorken, Ellis--Jaffe \dots and All That}, Univ. di Milano report 
(1993); more recently, see again: P.G.~Ratcliffe: {\sl Nuovo Cimento} {\bf A 
107} (1994) 2211, and references therein.}

\item{20.} {J.~Ellis and R.L.~Jaffe: {\sl Phys. Rev.} {\bf D 9} (1974) 1444; 
{\sl Phys. Rev.} {\bf D 10} (1974) 1669.}

\item{21.} {J.~Kodaira: {\sl Nucl. Phys.} {\bf B 165} (1980) 129.}

\item{22.} {G.~Altarelli and G.G.~Roos: {\sl Phys. Lett.} {\bf B 212} (1988) 
391; R.D.~Carlitz, J.C.~Collins and A.H.~Mueller: {\sl Phys. Lett.} {\bf B 
214} (1988) 229; G.~Altarelli and B.~Lampe: {\sl Z. Phys.} {\bf C 47} (1990) 
315. For the opposite view see: G.T.~Bodwin and J.~Qiu: {\sl Phys. Rev.} {\bf 
D 41} (1990) 2755. However the controversy seems to have been settled, 
hopefully in a definitive way, see: U.~Ellwanger: {\sl Phys. Lett.} {\bf B 
259} (1991) 469; W.~Vogelsang: {\sl Z. Phys.} {\bf C 50} (1991) 275; {\sl 
Nucl. Phys.} {\bf B 362} (1991) 3.}

\item{23.} {S.A.~Larin: {\sl Phys. Lett.} {\bf B 303} (1993) 113, 334.}

\item{24.} {Spin Muon Collaboration (D.~Adams, {\sl et al.}): {\sl Phys. 
Lett.} {\bf B 329} (1994) 399, {\sl erratum} {\bf B 339} (1994) 332.}

\item{25.} {BCDMS Collaboration (A.C.~Benvenuti, {\sl et al.}): {\sl Phys. 
Lett.} {\bf B 223} (1989) 485; {\sl Phys. Lett.} {\bf B 237} (1989) 592, 599.}

\item{26.} {New Muon Collaboration (P.~Amaudruz, {\sl et al}): {\sl Phys. 
Lett.} {\bf B 295} (1992) 159; {\sl Nucl. Phys.} {\bf B 371} (1992) 3.}

\item{27.} {G.~Altarelli, P.~Nason and G.~Ridolfi: {\sl Phys. Lett.} {\bf B 
320} (1994) 153, {\sl erratum} {\bf B 325} (1994) 538.}

\item{28.} {SLAC E143 Collaboration (K.~Abe, {\sl et al.}): report 
SLAC--PUB--95--6997 ({St}an\-ford, November 1995), also available as 
hep--ex/9511015. These data are a finer rebinning of those in 
Refs. 12--13, plus some data at higher $Q^2$ but over a restricted $x$--range, 
not included in the present analysis, completed {\sl before} they were 
available.}

\item{29.} {A.D.~Martin, W.J.~Stirling and R.G.~Roberts: {\sl Phys. Rev.} 
{\bf D 47} (1993) 867; {\sl Phys. Lett.} {\bf B 306} (1993) 145, {\sl 
erratum} {\bf B 309} (1993) 492; H.~Plothow--Besch: {\sl 
Comput. Phys. Commun.} {\bf 75} (1993) 396.}

\item{30.} {F.E.~Close: {\sl Perspectives in Nuclear Physics at Intermediate 
Energies, 6th Workshop}, ed. by S.~Boffi, C.~Ciofi~degli~Atti and 
M.M.~Giannini (World Scientific, Singapore 1994), p. 103, and Ref. 36.}

\item{31.} {See, for instance: J.~Bartelski and S.~Tatur: report CAMK 95--288 
(Warsaw, February 1995), also available as hep--ph/9502271.}

\item{32.} {G.~Altarelli, N.~Cabibbo, L.~Maiani and R.~Petronzio: {\sl Phys. 
Lett.} {\bf 48 B} (1974) 435, and the review by G.~Altarelli: {\sl Riv. Nuovo 
Cimento} {\bf 4} (1974) 335.}

\item{33.} {H.~Fritzsch: {\sl Phys. Lett.} {\bf B 229} (1989) 122; {\sl Mod. 
Phys. Lett.} {\bf A 5} (1990) 625, 1815; {\sl Phys. Lett.} {\bf B 256} (1991) 
75; report CERN--TH 7079--93 (Geneve, March 1994), also available as 
hep--ph/9403206.}

\item{34.} {S.J.~Brodsky and I.~Schmidt: {\sl Phys. Lett.} {\bf B 234} (1990) 
144; S.J.~Brodsky, M.~Burkardt and I.~Schmidt: {\sl Nucl. Phys.} {\bf B 441} 
(1995) 197.}

\item{35.} {The author recalls such parametrizations, which are scattered 
throughout the DIS literature, being presented as ``natural parametrizations'' 
in seminars on the parton model given by the late R.P.~Feynman at SLAC 
in the mid--seventies. For a paper reference, see: R.D.~Field and 
R.P.~Feynman: {\sl Phys. Rev.} {\bf B 15} (1977) 2590, where mentions of 
Pauli--principle effects can also be found.}

\item{36.} {An interesting reading, which the author finds as relevant to 
early QCD as the {\sl Blegdamsvej Faust} to quantum mechanics, is: 
A.~De~R\'ujula, J.~Ellis, R.~Petronzio, G.~Preparata and W.~Scott: report 
CERN--TH. 2778 (Geneve, November 1979). For an equally informative, but 
easier to find, reference see: F.E.~Close, {\sl An Introduction to 
Quarks and Partons} (Academic Press, London 1979).}

\item{37.} {The same conclusion have been independently reached and 
extensively check\-ed, though with different formulations, by: F.~Buccella, 
G.~Miele, G.~Migliore and V.~Tibullo: {\sl Z. Phys.} {\bf C 68} (1995) 631.}

\item{38.} {S.D.~Bass and P.V.~Landshoff: {\sl Phys. Lett.} {\bf B 336} (1994) 
537.}

\item{39.} {P.G.~Ratcliffe: {\sl Phys. Lett} {\bf B 242} (1990) 271, and Sez. 
I.N.F.N. di Milano report (September 1995), also available as 
hep--ph/9509237; M.~Roos: {\sl Phys. Lett.} {\bf B 246} (1990) 179; 
C.~Avenarius: {\sl Phys. Lett.} {\bf B 272} (1991) 71; B.~Ehrnsperger and 
A.~Sch\"afer: {\sl Phys. Lett.} {\bf B 348} (1995) 619.}

\item{40.} {P.M.~Gensini: {\sl Nuovo Cimento} {\bf A 103} (1990) 303; 
P.M.~Gensini and G.~Violini: $\pi N$ {\sl Newslett.} {\bf 9} (1993) 80, and 
Univ. di Perugia report, to appear soon.}

\item{41.} {R.L.~Jaffe and A.V.~Manohar: {\sl Nucl. Phys.} {\bf B 337} (1990) 
509.}

\item{42.} {Particle Data Group (L.~Montanet, {\sl et al.}): {\sl Phys. Rev.} 
{\bf D 50} (1994) 1173.}

\item{43.} {G.~Karl: {\sl Phys. Lett.} {\bf B 328} (1994) 149, {\sl 
erratum} {\bf B 322} (1994) 473. See also E.M.~Henley and G.A.~Miller: {\sl 
Phys. Rev.} {\bf D 50} (1994) 7077.}

\item{44.} {CERN--SPS Hyperon Beam Collaboration (M.~Bourquin, {\sl et al.}): 
{\sl Z. Phys.} {\bf C 12} (1982) 307.}

\item{45.} {S.~Hsueh, {\sl et al.}: {\sl Phys. Rev.} {\bf D 38} (1988) 2056.}

\item{46.} {Spin Muon Collaboration (D.~Adams, {\sl et al.}): {\sl Phys. 
Lett.} {\bf B 357} (1995) 248.}

\item{47.} {SLAC--Yale E80 Collaboration (M.J.~Alguard, {\sl et al.}): {\sl 
Phys. Rev. Lett.} {\bf 37} (1976) 1261; {\bf 41} (1978) 70; E130 Collaboration 
(G.~Baum, {\sl et al.}): {\sl Phys. Rev. Lett.} {\bf 45} (1980) 2000; {\bf 51} 
(1983) 1135.}

\item{48.} {The error corridor of the $\alpha_s(Q^2)$ determinations has 
a discontinuous momentum dependence, and it is advisable to use the error 
on this quantity rather than that on $\Lambda_{\overline{MS}}$ in 
error--propagation formul\ae, which is in practice what the above recipe 
is doing.}

\item{49.} {P.M.~Gensini: {\sl Nuovo Cimento} {\bf A 103} (1990) 1311; J.~Dai, 
R.F.~Dashen, E.~Jenkins, A.V.~Manohar: report UCSD--PTH--94--19 (San Diego, 
June 1995), also available as hep--ph/9506273.}

\item{50.} {F.E.~Close and R.G.~Roberts: {\sl Phys. Lett.} {\bf B 316} (1993) 
165; {\bf B 336} (1994) 257. See also F.E.~Close's summary talk presented at 
the 1995 Erice Conference, report RAL TR--95--047 (Chilton, September 1995), 
also available as hep--ph/9509251.}

\item{51.} {H.J.~Lipkin: {\sl Phys. Lett.} {\bf B 237} (1990) 130; {\bf B 251} 
(1990) 613; {\bf B 256} (1991) 284; {\bf B 337} (1994) 157.}

\item{52.} {Similar conclusions, based on different parametrizations and a 
study of the $x$--dependence, have been independently reached, although at 
next--to--leading order only, by F.~Buccella, O.~Pisanti, P.~Santorelli and 
J.~Soffer: report DSF--T--95/26 (Napoli, July 1995), also available as 
hep--ph/9507251.}

\bye